# On the Quantum Deviations from Einstein Dilation of Unstable Quanton* Decay Evolution and Lifetimes

by

Gordon N. Fleming
Professor Emeritus of Physics
Pennsylvania State University
Univ. Park, PA 16801

gnf1@earthlink.net          gnf1@psu.edu

**Abstract:** For over a decade several workers have argued for the existence of quantum deviations from the classical, Einstein dilation of the decay evolution of moving or Lorentz boosted unstable particles. While the general claim is correct, the discussions have been incomplete and, sometimes, misleading. The discussions have been of three kinds. Type 1 examines the time dependence of the survival probability for 3-momentum eigenstates of the unstable quanton (Khalfin). Type 2 does the same for velocity eigenstates, obtaining an outrageous result which then discredits velocity eigenstates (Shirokov / Hegerfeldt). Type 3 examines arbitrary boosts of 3-momentum eigenstates (Stefanovich). Type 1 is incomplete since the momentum eigenstates are not the boosts of one another. Type 2 is misleading since the outrageous result is due to misinterpreting the initial conditions of the velocity eigenstates (as I have previously argued). Type 3 is the most satisfactory, but has failed to recognize and implement the unification of all three types of discussion that can be achieved. In this paper I will provide that unified treatment, beginning with a recapitulation of Type 1 and offering further clarification of Type 2 in the process. The unified treatment fully reinstates velocity eigenstates as *essential* contributors to unstable quanton states. Besides discussing the time evolution of survival probabilities I also focus on the concept of lifetime defined as the *average time of decay*. This quantity is helpful in order to display the inequivalent dependence of dilation on momentum and boosts most sharply and the deviation from Einstein dilation most cleanly.

*I follow Jean-Marc Levy-Leblond's proposed terminology [Le 88, 99].

## 1. Introduction:

It is well known that if unstable quantons (UQs) are modeled within the framework of the standard quantum formalism, then the time evolution of such quantons in the form of exponential decay is, at best, an approximation [Ma 45; Hö 58; Kh 58; Pet 59a, b; Le 59; Ne 61; Fl 73; Fo 78; Pe 80]. Deviations occur at least at very long times due to the finite lower bound on the mass spectrum of the quanton and at short times if the initial state has a finite rms deviation of the energy spectrum or even a finite energy expectation value [Fo 78]. Another instance, but not so well known, of widespread use of what is only an approximation in the theory of unstable quanton decay is the classical, Einstein form of lifetime dilation.

In classical relativity the lifetime of an un-accelerated unstable particle is shortest relative to that inertial reference frame in which the particle is at rest. Relative to a boosted frame the lifetime is longer or 'dilated' as we customarily say. Equivalent to an account of how the lifetime varies with relatively moving inertial frames is an account of how the lifetime lengthens, or dilates, with increasing velocity or momentum of the particle. For a particle with mass, $m$, velocity, $\vec{v}$, or momentum, $\vec{k}$, the form taken by lifetime dilation is given by,

$$T_v = \frac{T_0}{\sqrt{1-v^2}} \ , \tag{1.1}$$

or,

$$T_k = T_0 \frac{\sqrt{m^2+k^2}}{m} \ , \tag{1.2}$$

respectively, where $T_0$ is the lifetime at zero velocity or momentum.

It has been shown, however, that in Lorentz covariant quantum theory the dependence of the decay evolution of an UQ on boost velocities and on momentum are, first of all, *not* equivalent to one another [St 96, 08] and second, are different in form, to one degree or another, from the classical case [Kh 97; St 96, 08; Sh 04]. These works focused on the time dependence of the survival probabilities rather than the lifetimes per se. While we will consider those time dependencies, our focus will be on the lifetimes which follow from them. Those lifetimes, defined in a manner independent of any presumptions about the form of the survival probabilities, will display the distinction between variation with momentum and boost velocities very sharply and the deviation from Einstein dilation very cleanly.

Briefly, the inequivalence between lifetime dilation due to momentum dependence and due to active boosting is a consequence of UQ 3-momentum eigenstates *not transforming into one another under Lorentz boosts*; which in turn is a consequence of the indefinite mass spectrum of the quanton. The deviation from Einstein dilation is also due to the indefinite mass spectrum for space-like momentum eigenstates and further due to the indefinite momentum spectrum for states of finite norm.

Besides the aforementioned studies there have also appeared claims [Sh 06, 09; He 06] to the effect that if one examines the survival probabilities of *velocity* eigenstates (which *do* transform into one another under Lorentz boosts) rather than 3-momentum eigenstates (they are *not* the same for UQs), the lifetimes *contract* with increasing velocity rather than dilate! This outrageous result then places in question the status, as basis vectors, of the velocity eigenstates. In my previous analysis of this work [Fl 09b] I argued that while the mathematics was correct, the interpretation of just what was being calculated was erroneous. Further clarification of that argument will be presented here. It turns out that the velocity eigenstates in question, far from constituting an alternative to the 3-momentum basis, are quite naturally seen as special cases of the generalized space-like momentum eigen-bases in the covariant formalism employed in [Fl 09b] and in this.

Practicing physicists may well ask why these issues have not arisen much earlier in the by now long history* of the theoretical and experimental study of unstable quanton decay. The answer lies in the empirical adequacy of simplifying approximations in the traditional approaches to decay; whether low order perturbation theory for metastable decays or Breit-Wigner resonance analysis for broad, unstable 'resonances'. In the former case one ignores the indefiniteness of the sharply peaked mass spectrum and in the latter one does not examine lifetimes because of their extreme brevity. Some have even suggested that the concept of lifetime is meaningless in the latter context [Lu 68].

---

*For a bibliography on some of that history see [Fo 78] and [Fl 09a].

The present paper, therefore, is an exercise in the careful analysis of these concepts and their relationships, predicated on the assumption of their joint meaningfulness, in principle, in *all* contexts notwithstanding the rarity of contexts in which they are all jointly and practically measureable, physically.

In section **2** we will declare and, perhaps, belabor somewhat our basic conceptual assumptions. These concern the space-time circumstances under which UQs can be isolated with unit probability and various aspects of their state vector representation. In section **3** we spell out the mathematical structure and time dependence of the survival probability for UQ states with definite **no-decay times**, defined in section **2**. In section **4** we will introduce the definition of lifetime that is logically prior to the exponential decay approximation, but which reduces to the standard usage if exponential decay is dominant. That prior definition is simply that the lifetime of an UQ is *the average time of decay* [Fl 73, 78]. Using that definition, we will, in sections **4** and **7**, come to see that the deviation of the quantum form of lifetime dilation from the classical Einstein form will be, after all, what we might have expected, intuitively, on quantum mechanical grounds. But a number of interesting and instructive surprises occur on the way to the final quantum form in **7**. Section **5** introduces the boosts of the previously considered states and develops a covariant formalism for their discussion. For these boosted states the no-decay times are replaced by **no-decay hyperplanes** which are not instantaneous. In section **6** we examine the survival probabilities for the general boosted states. We see that in the general case the space-like momentum eigenstate survival *probabilities* do not usually have corresponding survival *amplitudes* and display only kinematically trivial dependence on their "time" variables. Resolution of this puzzle is established only with the consideration of the generalized survival probability for finite norm states. Finally, in section **7** we obtain the generalized "lifetimes" for finding the boosted UQ states on arbitrary **survival hyperplanes**. The specialization of these "lifetimes" to *instantaneous* survival hyperplanes, albeit with non-instantaneous no-decay hyperplanes, yields the general *physical* lifetime for direct comparison with classical, Einstein dilation.

The results of the paper to which the author would particularly like to draw the readers attention are centered around the relations, (4.3, 4, 6), (6.4, 9, 15, 19) and (7. 5, 8, 10, 11).

To enhance the accessibility of this paper many calculational details, which the expert may regard as unnecessary, have been included in appendices.

Finally, for the comparison, in section **7**, of the general quantum lifetime dilation with the classical case it will be of use to consider the classical expression for the dilation, due to a boost velocity, $\vec{u}$, of the lifetime of an unstable particle already in motion with velocity, $\vec{v}$, or momentum, $\vec{k}$. The final velocity, $\vec{w}$, is compounded of $\vec{u}$ and $\vec{v}$ in the standard relativistic form,

$$\vec{w} = \frac{\vec{v}_\perp \sqrt{1-u^2}}{1+\vec{u}\cdot\vec{v}} + \frac{\vec{v}_\parallel + \vec{u}}{1+\vec{u}\cdot\vec{v}} \quad , \tag{1.3}$$

where the subscripts $\perp$ and $\parallel$ denote components perpendicular and parallel to $\vec{u}$, respectively. It then follows that,

$$\frac{1}{\sqrt{1-w^2}} = \frac{1+\vec{u}\cdot\vec{v}}{\sqrt{1-u^2}\sqrt{1-v^2}}, \tag{1.4}$$

from which we infer,

$$T_w = \frac{T_0}{\sqrt{1-w^2}} = \frac{T_v}{\sqrt{1-u^2}}[1+\vec{u}\cdot\vec{v}] = \frac{T_k}{\sqrt{1-u^2}}\left[1 + \frac{\vec{u}\cdot\vec{k}}{\sqrt{m^2+k^2}}\right], \tag{1.5}$$

where, $T_v = T_k$, is the lifetime of the particle with velocity, $\vec{v}$, or momentum, $\vec{k}$. As we will see in **7**, the algebraic form of (1.5) acquires enhanced significance in the quantum domain.

Throughout the paper quantum mechanical operators are identified by a circumflex as in $\hat{P}^\mu$.

## 2. Quantons undecayed at definite times

UQs are distinguished both from unstable classical particles and from stable quantons. The modes of distinction are different in each case and both are important.

The quantum states of isolated stable quantons are identified as being discrete spectrum eigenstates of two Casimir invariants of the Poincare space-time symmetry group, the squared rest mass and the squared internal angular momentum, or spin. For a given quanton both eigenvalues are fixed throughout the range of its possible states. While an UQ can not have a definite rest mass, we will here keep the distinction between stable and unstable quantons at a minimum by retaining a definite spin* and having the rest mass spectral function for a given UQ type be independent of the varied quantum states which the quanton type can assume. These conditions, which are employed by Khalfin, Stefanovich, Shirokov and Hegerfeldt will be implemented explicitly below.

Unstable classical particles, being *particles*, have their existence confined to or within a small neighborhood of a world line of finite extent. Every point of that world line has a definite time coordinate in every inertial frame. In particular, the *initial point* at which the unstable particle has unit probability of being undecayed corresponds to a definite time in each inertial frame.

Quantons, however, stable or unstable, do not have their existence confined to the neighborhoods of worldlines. Unit norm states are of necessity, and in position representation explicitly, space-like extended and with infinite tails in the more realistic examples. Momentum eigenstates and normed approximate momentum eigenstates, which play such an important role in our analyses of dynamical processes, are emphatic examples of the space-like extension of quanton states. Consequently, the analogue, for quantons, to particles existing at space-time points of a world line, will here be taken to be existence on space-like hyperplanes. For stable quantons, isolation on one hyperplane, i.e., at one time in one inertial frame, entails isolation at any time in any inertial frame, i.e., on all hyperplanes. For an UQ, however, being isolated, i.e., being alone and undecayed with unit probability, can hold at only one instant of time in some inertial frame, i.e., on only one space-like hyperplane.

Working in the Heisenberg picture, we begin with a pure state for a system which, at the time, $t = 0$, consists, along with the vacuum, of one undecayed, UQ in the momentary single quanton state labeled by $\psi$. At any other time the probability for finding **decay products** or **formation**

*Relaxations of the spin constraint have been considered in [Be 62] and [Fl 72, 79].

**precursors** instead of the UQ will be greater than zero. That unique time when the probability for finding the UQ is unity will be called the **no-decay time**. The unit norm, Heisenberg picture state vector for this system will be denoted by, $|\psi; 0\rangle$, or, for brevity, $|\psi\rangle$.

Upon applying the unitary operators representing Euclidean transformations and time translations to this state vector we will write,

$$\hat{U}(b)\hat{U}(\bar{a})\hat{U}(R)|\psi;0\rangle = |\psi_{R,\bar{a}}; b\rangle, \qquad (2.1)$$

where $R$ denotes the rotation and, $\bar{a}$, the spatial translation of the Euclidean transformation and $b$ denotes the time translation. In the resulting state, $|\psi_{R,\bar{a}}; b\rangle$, the time translation parameter, $b$, denotes the new no-decay time at which the UQ is certain to be found undecayed. The momentary state for the quanton that holds at the time, $t = b$, is the Euclidean transformed state, $\psi_{R,\bar{a}}$. If $b \neq 0$, then at $t = b$ in the original state, $|\psi;0\rangle$, decay products (b > 0) or formation precursors (b < 0) can be found with non-zero probability and the probability for finding the parent quanton is less than unity. Similarly, in the transformed state, $|\psi_{R,\bar{a}}; b\rangle$, the probabilities for finding decay products/formation precursors or the single parent at $t = 0$ are greater than zero and less than unity, respectively. Unlike the time translation, which has these dynamical consequences, the Euclidean transformation is dynamically innocuous in the sense that it leaves the no-decay time unchanged and only modifies the momentary state of the undecayed parent. When, in section **5**, we consider Lorentz boosts of UQ states with no-decay times, we will be dealing with UQ states that do not have no-decay *times* but, instead, *non-instantaneous, space-like* **no-decay hyperplanes**. This motivates distinguishing our present subclass of UQ states by the name of **Instantaneous Single Parent** (ISP) states. For each no-decay time, $t$, the ISP states for a single *type* of UQ comprise a linear state space that is invariant under the Euclidean group.

One consequence of the dynamical innocuousness of spatial translations for ISP states is that the latter include and are spanned by 3-momentum eigenstates. The construction of such eigenstates entails only the superposition of spatial translations of a single state with a definite no-decay time. Thus we have,

$$| \psi, \vec{k}; t > = \delta^3(\hat{\vec{P}} - \vec{k}) | \psi; t > = (2\pi\hbar)^{-3} \int d^3\lambda \, \exp[(i/\hbar)(\hat{\vec{P}} - \vec{k}) \cdot \vec{\lambda}] | \psi; t > . \qquad (2.2)$$

Since,

$$| \psi; t > = \int d^3k \, | \psi, \vec{k}; t > , \qquad (2.3)$$

the result, $| \psi, \vec{k}; t >$, must be non-vanishing for some $\vec{k}$.

On the other hand, the ISP states can not include any *energy* eigenstates at all, since the spontaneous decay evolution that defines an UQ demands, via the energy-time uncertainty relation, an energy spectrum of non-zero width. That spectrum stems from the rest-mass spectrum defined by the rest-mass spectral function,

$$\sigma_\psi(\mu) := 2\mu < \psi; t \, | \, \delta(\hat{P}^2 - \mu^2) \, | \, \psi; t > / \, \| \, \psi \, \|^2 \, , \qquad (2.4)$$

independent of $t$ due to $\hat{P}^0$ commuting with $\hat{P}^2 = (\hat{P}^0)^2 - (\hat{\vec{P}})^2$. But what of the dependence on the state, $\psi$? As indicated above we minimize the deviation from stability by following the lead of the works we are commenting on and *assuming* the invariance over the state space of the mass spectral function, i.e..

$$\sigma_\psi(\mu) = \sigma(\mu) . \qquad (2.5)$$

We will explicitly extend (2.5) to the 3-momentum eigenstates below. We might well have *derived* this result from a specification of how the various 3-momentum eigenstates are related to one another. If the transformation between such eigenstates (which can not be, as mentioned above, a Lorentz boost, due to the indefinite energy spectrum) commutes with the invariant mass operator, the state independence of the mass spectral function follows.

But first we must comment on the *degeneracy* of the 3-momentum eigenvalues. In accordance with our opening assumption we will have a degeneracy of $2s + 1$ for each eigenvalue due to the UQ having a definite spin, *s*. Thus we have a translationally invariant, 3-vector operator, $\hat{\vec{S}}$, such that,

$$\hat{\vec{S}}^2 | \psi, \vec{k}; t > = | \psi, \vec{k}; t > \hbar^2 s(s+1) \, , \qquad (2.6)$$

for arbitrary $\psi$ and $\vec{k}$, where ($\hat{\vec{J}}$ is the total angular momentum and generator of rotations),

$$[\hat{J}^m, \hat{S}^n] = [\hat{S}^m, \hat{S}^n] = i\hbar\, \varepsilon^{mnl}\, \hat{S}^l \ . \qquad (2.7)$$

and which is otherwise related to the ISP states with a given no-decay time in exactly the same way it would relate to the states of a stable quanton with spin $s$ (For details see **App. 1**).

From (2.6, 7) it follows that we can write,

$$|\psi, \vec{k}; t> = \sum_{m=-s}^{s} |\vec{k}, m; t> \psi_m(\vec{k}) \ , \qquad (2.8)$$

where, with appropriately chosen Cartesian axes, we have,

$$\hat{S}^3 |\vec{k}, m; t> = |\vec{k}, m; t> \hbar m \ . \qquad (2.9)$$

If we normalize these basis vectors according to,

$$<\vec{k}', m'; t\, |\, \vec{k}, m; t> = \delta^3(\vec{k}' - \vec{k})\delta_{m'm} \ , \qquad (2.10)$$

then, from (2.3, 4, 5, 8), we have,

$$2\mu <\vec{k}', m'; t\, |\, \delta(\hat{P}^2 - \mu^2)\, |\, \vec{k}, m; t> = \delta^3(\vec{k}' - \vec{k})\delta_{m'm}\, \sigma(\mu) \ . \qquad (2.11)$$

All the equations, (2.8-11), leave a phase factor undetermined in the definition of the momentum-spin eigenstates, $|\vec{k}, m; t>$. The phase factor can always be chosen so that,

$$\hat{\vec{J}} |\vec{k}, m; t> = i\hbar\left(\vec{k} \times \frac{\partial}{\partial\vec{k}}\right) |\vec{k}, m; t> + \hat{\vec{S}} |\vec{k}, m; t> \ . \qquad (2.12a)$$

With that choice, then, we can, in the ISP state space, $Span_{\vec{k}, m}\{|\vec{k}, m; t>\}$, employ an operator, $\hat{\vec{X}}(t)$, defined by,

$$\hat{\vec{X}}(t) |\vec{k}, m; t> = -i\hbar \frac{\partial}{\partial\vec{k}} |\vec{k}, m; t> \ , \qquad (2.12b)$$

to obtain (**App. 1**),

$$\hat{\vec{J}} = \hat{\vec{X}}(t) \times \hat{\vec{P}} + \hat{\vec{S}} \, . \tag{2.13}$$

After we turn to a consideration of Lorentz boosts in section **5** we will be able to complete the definition of this position operator without the restriction to the single quanton subspace. We will then recognize it as the generalized Newton-Wigner position operator [New 49] which, because of its relation to spin in the correspondingly generalized version of (2.13), can equally well be characterized as *the center of spin* position operator [Fl 99].

Finally, the projector onto the state space for ISP states with no-decay time, *t*, is given by,

$$\hat{\Pi}(t) = \sum_{m=-s}^{s} \int d^3k \, | \, \vec{k}, m; t > < \vec{k}, m; t \, | \, . \tag{2.14}$$

## 3. The survival probability for ISP states

The 3-momentum eigenstates, $| \, \vec{k}, m; t >$, are, themselves, linear superpositions of 4-momentum eigenstates, $| \, q, m >$. Explicitly we have [St 08] (**App. 2**),

$$| \, \vec{k}, m; t > = \int d^4q \, | \, q, m > \delta^3(\vec{q} - \vec{k}) \sqrt{2q^0} \, r(q) \exp[(i / \hbar) q^0 t] \, , \tag{3.1}$$

where,

$$| \, r(q) \, |^2 = \sigma\left(\sqrt{q^2}\right) / 2\sqrt{q^2} \, , \tag{3.2}$$

and,

$$< q', m' \, | \, q, m > = \delta^4(q' - q) \delta_{m'm} \, . \tag{3.3}$$

A more detailed examination of the 4-momentum-spin eigenstates, $| \, q, m >$, would find them to contain the information concerning the number, type and angular distribution of the decay products of the UQ in the asymptotic future or the formation precursors of the UQ in the asymptotic past. As our focus in this paper is on the time-like evolution of the UQ survival probability and the momentum and boost dependence of the UQ lifetime, we will not examine the deeper structure of the $| \, q, m >$ basis states here.

We are now in the position to examine the time dependence of the **survival probability**, $P_k(t)$, for an ISP 3-momentum eigenstate. The definition of that probability is given by,

$$< \vec{k}', m'; 0 \mid \hat{\Pi}(t) \mid \vec{k}, m; 0 > = \delta^3(\vec{k}' - \vec{k}) \delta_{m'm} P_k(t) , \tag{3.4}$$

where the rotational invariance of the projector, $\hat{\Pi}(t)$, (2.14), guarantees that the survival probability will be independent of the direction of the 3-momentum or the value of the spin component number, $m$.

Because of the conservation of 3-momentum through time, only an infinitesimal range of the momentum integral that defines the projector, (2.14), contributes to the survival probability and, consequently, that probability is the absolute square of the survival *amplitude*, $I_k(t)$, defined by,

$$< \vec{k}', m'; t \mid \vec{k}, m; 0 > = \delta^3(\vec{k}' - \vec{k}) \delta_{m'm} I_k(t) . \tag{3.5}$$

The only reason for drawing attention to this relation between probability and amplitude, which is so ubiquitous throughout quantum theory, is that we will see instances where it does not hold when we turn to consider Lorentz boosted UQ states.

From (3.1-3, 5) the survival amplitude is given by,

$$I_k(t) = \int_{\mu_{\min}}^{\infty} d\mu \, \sigma(\mu) \exp\left[-(i/\hbar)\sqrt{\mu^2 + k^2} \; t\right] , \tag{3.6}$$

and the survival probability by,

$$P_k(t) = \left| \int_{\mu_{\min}}^{\infty} d\mu \, \sigma(\mu) \exp\left[-(i/\hbar)\sqrt{\mu^2 + k^2} \; t\right] \right|^2$$

$$= \int_{\mu_{\min}}^{\infty} d\mu' \int_{\mu_{\min}}^{\infty} d\mu \, \sigma(\mu') \sigma(\mu) \exp\left[(i/\hbar)\left(\sqrt{\mu'^2 + k^2} - \sqrt{\mu^2 + k^2}\right)t\right]$$

$$= 2 \int_{\mu_{\min}}^{\infty} d\mu' \, \sigma(\mu') \int_{\mu_{\min}}^{\mu'} d\mu \, \sigma(\mu) \cos\left[\left(\int_{\mu}^{\mu'} d\mu'' \frac{\mu''}{\sqrt{\mu''^2 + k^2}}\right)\frac{t}{\hbar}\right] . \tag{3.7}$$

In the last form of (3.7) we most easily discern the influence of the factor,

$$\frac{\mu''}{\sqrt{\mu''^2 + k^2}} \ , \tag{3.8}$$

which slows down the evolution as $k$, the magnitude of momentum, increases. But in this instance the slowdown, although of the intuitively expected form, at least on average, and leading to lifetime dilation, has no *immediate* connection with Lorentz boosts. The increase in the value of $k$ does not arise by Lorentz boosting the UQ state from $k = 0$, but by a unitary transform generated, via (2.12b), by the position operator,

$$|\vec{k}, m; t> = \exp[(i/\hbar)\hat{\vec{X}}(t) \cdot \vec{k}] \, | \vec{0}, m; t> . \tag{3.9}$$

Of course the *form* of the dependence on momentum of the slowing of decay comes from the *form* of the dependence of the energy on the momentum for the mass eigenstates, $|q, m>$, that contribute to the UQ state. Those mass eigenstates *do* arise by boosting from $\vec{q} = \vec{0}$. But *those* boosts, in order to contribute to a given 3-momentum, $\vec{k}$, require different velocities for different mass eigenvalues and do not coalesce into any single boost for the UQ state as a whole.

For a unit norm ISP state, (2.8), with no-decay time, $t = 0$, the survival probability is given by,

$$P_\psi(t) = <\psi; 0 \,|\, \hat{\Pi}(t) \,|\, \psi; 0> = \int d^3k \left( \sum_{m=-s}^{s} |\psi_m(\vec{k})|^2 \right) P_k(t), \tag{3.10}$$

and here the deviation from classical, Einstein retardation of the time evolution results from integration over both the indefinite mass spectrum and the indefinite momentum spectrum. In effect, these are the deviations reported by Khalfin [Kh 97], Stefanovich [St 96, 08] and Shirokov [Sh 04].

## 4. The lifetime for ISP states:

The survival probabilities (3.4, 7) and (3.10) can approximate exponential decay rather well over a wide range of circumstances (however, see [Ke 10]), but they can never be exactly exponential in form. As indicated by (3.7) the exact time dependence is determined by the mass spectral function, (2.4, 11), and any deviation in that from the Breit-Wigner resonance formula with unconstrained support (which deviations are always present) yields deviations from exponential decay. In fact, as Khalfin showed [Kh 97] even if the mass spectral function were exactly Breit Wigner in form and $P_0(t)$ were exactly exponential, it follows from (3.7) that $P_k(t)$, for $k > 0$, would not be so and, from (3.10), that $P_\psi(t)$ would not be so. Nevertheless, we are so accustomed to associating the concept of lifetime with exponential decay that we tend to think of the definition of lifetime as that parameter in the denominator of the exponent of the exponential decay form, $\exp[-t/T]$. But, as mentioned in the introduction, the *definition* of lifetime is simply **the average time of decay**.

The probability for decay to occur in the time interval between $t$ and $t + dt$ is taken to be the product of the rate of decay and $dt$. For 3-momentum eigenstates this is just, $-\dot{P}_k(t)dt$, except that quantum mechanically, $P_k(t)$, need not be monotonically decreasing with time. It can, occasionally, display regeneration and briefly increase with time, thereby briefly yielding a negative "rate of decay". Nevertheless, we will employ the standard expression for the average time of decay [Fl 73, 78; Fo 78],

$$T_k = -\int_0^\infty t\,\dot{P}_k(t)\,dt\,, \qquad (4.1a)$$

to define the lifetime, $T_k$. Integration by parts converts this into,

$$T_k = \int_0^\infty P_k(t)\,dt\,. \qquad (4.1b)$$

If the survival probability is dominated by the exponential form, then (4.1b) will yield a result correspondingly close to the parameter in the denominator of the exponent. An awkwardness with this lifetime formula is that the integral can diverge to infinity if the survival probability does not vanish asymptotically more rapidly than $t^{-1}$. But it has long been known [Lé 59] that if the threshold dependence of the spectral function satisfies,

$$\sigma(\mu) \approx (\mu - \mu_{\min})^\alpha \qquad (4.2a)$$

with $\alpha > 0$ then,

$$P_k(t) \approx t^{-2(1+\alpha)} \qquad (4.2b)$$

asymptotically as $t \to \infty$.

If we now substitute (3.7) for $P_k(t)$ we find (**App. 3**),

$$T_k = \int\limits_0^\infty dt \left| \int\limits_{\mu_{\min}}^\infty d\mu \, \sigma(\mu) \exp\left[ -(i/\hbar)\sqrt{\mu^2 + k^2}\, t \right] \right|^2$$

$$= \pi\hbar \int\limits_{\mu_{\min}}^\infty d\mu \, \sigma(\mu)^2 \left( \sqrt{\mu^2 + k^2} / \mu \right) . \tag{4.3}$$

We immediately see the dilation of the lifetime with increasing momentum in accordance with the relativistic energy factor, $\sqrt{\mu^2 + k^2} / \mu$, but as with the survival probability (3.7) and unlike the classical case, this energy factor is integrated over all the contributing mass values represented in the mass spectral function. The lifetime is *not* dilated via one multiplicative energy factor for some definite mass as in (1.2). Furthermore, as mentioned above after (3.8), we have yet to consider any Lorentz boost of an UQ state! The contributing relativistic energy factors do not arise from a single Lorentz transform but from the family of Lorentz transforms correlated to each contributing mass and tailored to boost that mass contribution to the same 3-momentum, $\vec{k}$.

If we now consider a unit norm ISP state, $| \psi; 0 >$, with no-decay time, *t = 0*, and survival probability given by (3.10) we find the lifetime,

$$T_\psi = \int\limits_0^\infty dt \, P_\psi(t) = \sum_{m=-s}^s \int d^3k \, | \psi_m(\vec{k}) |^2 \left[ \pi\hbar \int\limits_{\mu_{\min}}^\infty d\mu \, \sigma(\mu)^2 \left( \sqrt{\mu^2 + k^2} / \mu \right) \right] , \tag{4.4}$$

with a lifetime dilation factor integrated over all contributing 3-momenta as well as all contributing masses. Needless to say, these deviations, aside from being quantitatively small in practice, in no sense constitute a conflict with Lorentz covariance, but are exactly to be expected from quantum mechanical states of indefinite mass and 3-momentum.

By considering (2.4, 5, 11) and the operator identity,

$$\int d\mu \, 2\mu \delta(\mu^2 - \hat{P}^2) \sigma(\mu) \frac{\sqrt{\mu^2 + \hat{\vec{P}}^2}}{\mu} = \sigma\left( \sqrt{\hat{P}^2} \right) \frac{\hat{P}^0}{\sqrt{\hat{P}^2}} , \tag{4.5}$$

we obtain the following compact expression for the lifetime (4.4),

$$T_\psi = \pi\hbar < \psi \,|\, \sigma\!\left(\sqrt{\hat{P}^2}\right) \frac{\hat{P}^0}{\sqrt{\hat{P}^2}} \,|\, \psi > .$$
(4.6)

## 5. Boosting ISP states:

We now turn to the examination of UQ states obtained by Lorentz boosting ISP states, i.e., states of the form, $\hat{U}(B)|\psi;t>$, where $B$ is a Lorentz boost. We begin with the consideration of boosts of 3-momentum-spin eigenstates.

From the principle of Lorentz covariance such boosted states must present the same features as the original ISP, 3-momentum-spin eigenstate would present to an inertial frame, $F'$, obtained via the inverse boost. Relative to such an inertial frame the original state would not be a 3-momentum eigenstate, but would rather be an eigenstate of appropriately defined space-like momenta. The state would also not have any no-decay *time* in $F'$, i.e., a definite time for which the probability for finding the undecayed parent quanton is unity. Instead, the no-decay time would be replaced by a ***non-instantaneous* no-decay hyperplane**; namely that same hyperplane which was labeled by the no-decay time in the original frame. For an active boost of an ISP state (with no-decay time, $t$ ) by velocity, $\vec{u}$, $B(\vec{u})$, the no-decay hyperplane will be orthogonal to the time-like unit 4-vector,

$$\eta^\mu = (\eta^0, \vec{\eta}) = B(\vec{u})(1,\vec{0}) = \left( \frac{1}{\sqrt{1-u^2}}, \frac{\vec{u}}{\sqrt{1-u^2}} \right),$$
(5.1)

and will contain the point with Minkowski coordinates, $x^\mu = \eta^\mu \tau = \eta^\mu ct$. We will label such hyperplanes with the ordered pair, $(\eta, \tau)$ .The space-like momentum eigenvalue for the boosted state will be, $p^\mu$, where,

$$p^\mu = (p^0, \vec{p}) = B(\vec{u})(0,\vec{k}) = \left( \frac{\vec{u}\cdot\vec{k}}{\sqrt{1-u^2}}, \vec{k}_\perp + \frac{\vec{k}_\parallel}{\sqrt{1-u^2}} \right),$$
(5.2)

(the subscripts $\perp$ and $\parallel$ denoting components orthogonal and parallel, respectively, to the boost velocity) satisfying, $\eta p = 0$ and $p^2 = -\vec{k}^2$. Thus we can write,

$$\hat{U}(B(\vec{u}))|\,\vec{k},m;t> = |\,p,m;\eta,\tau=ct>,$$
(5.3)

where,

$$(\hat{P}^{\mu} - \eta^{\mu}(\eta\hat{P}))\,|\,p,m;\eta,\tau> = |\,p,m;\eta,\tau> p^{\mu}\,. \qquad (5.4)$$

The boosted spin eigenvalue, $m$, is to be understood as follows: From,

$$\hat{U}^{\dagger}(B(\vec{u}))\hat{S}^{\mu}(\eta)\hat{U}(B(\vec{u})) = \hat{U}^{\dagger}(B(\vec{u}))(\hat{S}^0(\eta),\hat{\vec{S}}(\eta))\hat{U}(B(\vec{u}))\,,$$

$$= B(\vec{u})(\hat{S}^0((1,\vec{0})),\hat{\vec{S}}((1,\vec{0}))) = B(\vec{u})(0,\hat{\vec{S}}) = \left(\frac{\vec{u}\cdot\hat{\vec{S}}}{\sqrt{1-u^2}},\hat{\vec{S}}_{\perp} + \frac{\hat{\vec{S}}_{\parallel}}{\sqrt{1-u^2}}\right)\,, \qquad (5.5)$$

it follows that (**App. 4**),

$$\left[\hat{\vec{S}}(\eta) - \frac{\vec{\eta}}{\eta^0+1}\hat{S}^0(\eta)\right]\hat{U}(B(\vec{u})) = \hat{U}(B(\vec{u}))\,\hat{\vec{S}}((1,\vec{0})) = \hat{U}(B(\vec{u}))\,\hat{\vec{S}}\,. \qquad (5.6)$$

Consequently, from (2.9), (5.3) and (5.6), we have,

$$\left[\hat{S}^3(\eta) - \frac{\eta^3}{\eta^0+1}\hat{S}^0(\eta)\right]|\,p,m;\eta,\tau> = |\,p,m;\eta,\tau> \hbar\,m \quad. \qquad (5.7)$$

For a general homogeneous Lorentz transformation, $\Lambda$, we must have,

$$\hat{U}(\Lambda)\,|\,p,m;\eta,\tau> = \sum_{m'=-s}^{s}|\,\Lambda p,m';\Lambda\eta,\tau> Y_{m',m}^s(R)\,, \qquad (5.8)$$

where,

$$R = B^{-1}(\vec{\eta}'/\eta'^0)\Lambda B(\vec{\eta}/\eta^0)\,, \qquad (5.9)$$

is, for $\eta' = \Lambda\eta$, a rotation (since, from (5.9), $R(1,\vec{0}) = (1,\vec{0})$ ) and the $Y_{m',m}^s(R)$ are spherical harmonic functions.

Accordingly, for unit norm states we have,

$$|\psi_B;\eta,\tau> = \hat{U}(B(\vec{u}))\,|\psi;t> = \sum_{m=-s}^{s}\int d^3k\,\hat{U}(B(\vec{u}))\,|\,\vec{k},m;t> \psi_m(\vec{k})$$

$$= \sum_{m=-s}^{s}\int d_{\eta}^3 p\,|\,p,m;\eta,\tau> \psi_{B,m}(p)\,, \qquad (5.10)$$

where,

$$d^3_\eta p = d^4 p \, \delta(\eta p) = d^4 k \, \delta(k^0) = d^3 k \quad , \tag{5.11a}$$

and

$$\psi_{B,m}(p) = \psi_m(B^{-1}p) = \psi_m((0, \vec{k})) = \psi_m(\vec{k}) \, , \tag{5.11b}$$

and, finally,

$$\hat{U}(\Lambda) | \psi; \eta, \tau > = | \psi_\Lambda; \Lambda\eta, \tau > , \tag{5.12}$$

where,

$$\psi_{\Lambda, m'}(\Lambda p) = \sum_{m=-s}^{s} Y_{m', m}^s(R) \, \psi_m(p) \, , \tag{5.13}$$

and $R$ is given by (5.9).

These space-like momentum - spin eigenstates and all useful superpositions of them, with fixed no-decay hyperplane in the superposition, comprise, along with the ISP states of the preceding sections, the possible states for an isolated UQ with some definite no-decay hyperplane. We will designate them, collectively, as **single parent** states (SP).

Having introduced Lorentz transforms, we can now complete the earlier preliminary remarks about the transformations between momentum eigenstates generated by the generalized Newton-Wigner position operator.

Writing, $\Lambda^\mu_\nu = \left(e^\omega\right)^\mu_\nu$, we have,

$$\hat{U}(\Lambda) = \exp[-(i/2\hbar)\hat{M}^{\mu\nu}\omega_{\mu\nu}]. \tag{5.14}$$

The generator of homogeneous transformations, $\hat{M}^{\mu\nu}$, can be decomposed, relative to any future pointing time-like unit vector, $\eta^\mu$, as

$$\hat{M}^{\mu\nu} = \hat{N}^\mu(\eta)\eta^\nu - \hat{N}^\nu(\eta)\eta^\mu + \varepsilon^{\mu\nu\alpha\beta}\hat{J}_\alpha(\eta)\eta_\beta \, , \tag{5.15}$$

where, $\eta\hat{N}(\eta) = \eta\hat{J}(\eta) = 0$. The generalized spin and NW position operator, $\hat{S}^\mu(\eta)$ and $\hat{X}^\mu(\eta, \tau)$, respectively, for arbitrary closed systems are then implicitly defined by the equations [Fl 99],

$$\hat{J}^\mu(\eta) = -\varepsilon^{\mu\alpha\beta\gamma}\hat{X}_\alpha(\eta, \tau)\hat{P}_\beta \eta_\gamma + \hat{S}^\mu(\eta) \, , \tag{5.16a}$$

and,

$$\hat{N}^\mu(\eta) = \eta \hat{P} : \hat{X}^\mu(\eta, \tau) - \hat{P}^\mu \tau - \frac{\varepsilon^{\mu\alpha\beta\gamma} \hat{P}_\alpha \hat{S}_\beta(\eta) \eta_\gamma}{\eta \hat{P} + \sqrt{\hat{P}^2}} , \qquad (5.16b)$$

where the colon indicates a symmetrized product. From these definitions, (2.12b) and (3.9) we can consistently assume, (**App. 5**)

$$\exp[-(i/\hbar)\hat{X}^\mu(\eta, \tau)\Delta_\mu] \, | \, p, m; \eta, \tau > = | \, p + \Delta, m; \eta, \tau > , \qquad (5.17)$$

where, $\eta\Delta = 0$.

# 6. Survival probability for SP states:

We now turn to the analysis of the survival probability for boosts of ISP states, which, as the preceding discussion makes clear, are, themselves, SP states with non-instantaneous no-decay hyperplanes. The quantities of interest are,

$$< \psi; 0 \, | \, \hat{U}^\dagger(B(\vec{u}))\hat{\Pi}(t)\hat{U}(B(\vec{u})) \, | \, \psi; 0 >$$

$$= < \psi_B; \eta, 0 \, | \, \hat{\Pi}(t) \, | \, \psi_B; \eta, 0 > = P_{\psi_B}(t \, | \, \eta, 0) \qquad (6.1)$$

for unit norm SP states, and,

$$< \vec{k}_2, m_2; 0 \, | \, \hat{U}^\dagger(B(\vec{u}))\hat{\Pi}(t)\hat{U}(B(\vec{u})) \, | \, \vec{k}_1, m_1; 0 >$$

$$= < p_2, m_2; \eta, 0 \, | \, \hat{\Pi}(t) \, | \, p_1, m_1; \eta, 0 > \propto P_{(p_2, m_2; p_1, m_1)}(t \, | \, \eta, 0) \qquad (6.2)$$

for SP momentum-spin eigenstates, where, as before, $B(\vec{u})(1, \vec{0}) = B(\vec{u})\eta^{(0)} = \eta$, and $B(\vec{u})(0, \vec{k}_{1,2}) = p_{1,2}$. The proportionality sign in the last stage of (6.2) is due to delicacies concerning momentum conserving delta functions. A natural question to ask at this point is whether, as was the case for ISP momentum eigenstates, the survival probability, $P_{(p_2, m_2; p_1, m_1)}(t \, | \, \eta, 0)$, is, itself, the absolute square of a survival amplitude, $I_{B(\vec{u})(0, \vec{k}_1)}(t)$, defined by,

$$< \vec{k}_2, m_2; 0 \, | \, \hat{U}^\dagger(B(\vec{u}))\exp[(i/\hbar)\hat{P}^0 t]\hat{U}(B(\vec{u})) \, | \, \vec{k}_1, m_1; 0 > = \delta^3(\vec{k}_2 - \vec{k}_1) I_{B(\vec{u})(0, \vec{k}_1)}(t) , \qquad (6.3)$$

where the delta function is, here, unquestioned. The answer is no, such probabilities *do not have corresponding amplitudes*, precisely because the matrix element (6.2), as we will see below, is not proportional to a *three* dimensional delta function, i.e., it does not conserve all the momentum components and, therefore, more than one term from the projection operator contributes to the probability.

But it is worth pausing here to examine a side issue and see what results if we blithely proceed as if (6.3) *did* define the survival *amplitude* for the boosted momentum eigenstate. We then find,

$$< \vec{k}_2, m_2; 0 \,|\, \hat{U}^\dagger(B(\vec{u})) \exp[(i/\hbar)\hat{P}^0 t] \hat{U}(B(\vec{u})) \,|\, \vec{k}_1, m_1; 0 >$$

$$= < \vec{k}_2, m_2; 0 \,|\, \hat{U}^\dagger(B(\vec{u})) \hat{U}(B(\vec{u})) \exp\left[ (i/\hbar) \frac{\hat{P}^0 + \vec{u} \cdot \hat{\vec{P}}}{\sqrt{1-u^2}} \, t \right] |\, \vec{k}_1, m_1; 0 >$$

$$= < \vec{k}_2, m_2; 0 \,|\, \exp\left[ (i/\hbar) \frac{\hat{P}^0 + \vec{u} \cdot \hat{\vec{P}}}{\sqrt{1-u^2}} \, t \right] |\, \vec{k}_1, m_1; 0 >$$

$$= < \vec{k}_2, m_2; 0 \,|\, \exp\left[ (i/\hbar) \frac{\hat{P}^0}{\sqrt{1-u^2}} \, t \right] |\, \vec{k}_1, m_1; 0 > \exp\left[ (i/\hbar) \frac{\vec{u} \cdot \vec{k}_1}{\sqrt{1-u^2}} \, t \right]. \qquad (6.4a)$$

Equating absolute values of the beginning and last expressions, we find,

$$\left| I_{B(\vec{u})(0, \vec{k}_1)}(t) \right| = \left| I_{k_1}\left( \frac{t}{\sqrt{1-u^2}} \right) \right|, \qquad (6.4b)$$

*and the boosted amplitude decays faster than the original amplitude!* This is, essentially, a generalization of the calculation Shirokov [Sh 06, 09] and Hegerfeldt [He 06] reported except that they confined themselves to the case, $\vec{k}_1 = \vec{0}$, because they were working with velocity eigenstates,

$$\frac{\hat{\vec{P}}}{\hat{P}^0} |\, \vec{u} > = \frac{\hat{\vec{P}}}{\hat{P}^0} \hat{U}(B(\vec{u})) |\, \vec{k} = \vec{0} > = \hat{U}(B(\vec{u})) \left[ \frac{\hat{\vec{P}}_\perp + (\hat{\vec{P}}_\parallel + \vec{u}\hat{P}^0)/\sqrt{1-u^2}}{(\hat{P}^0 + \vec{u} \cdot \hat{\vec{P}})/\sqrt{1-u^2}} \right] |\, \vec{k} = \vec{0} >$$

$$= \hat{U}(B(\vec{u})) \frac{\vec{u}\hat{P}^0}{\hat{P}^0} |\, \vec{k} = \vec{0} > = \hat{U}(B(\vec{u})) |\, \vec{k} = \vec{0} > \vec{u} = |\, \vec{u} > \vec{u}. \qquad (6.5)$$

Unlike 3-momentum eigenstates, *velocity eigenstates do transform into one another under Lorentz boosts* and this has made them attractive to various workers [Zw 63; Ha 72; Ra 73; Se 75; Bo 00; Ta 08] as possible UQ basis vectors. But (6.4) is clearly a problem and so Shirokov put them aside and returned to momentum eigenstates. But this leaves the question of the physical significance of (6.4) hanging and I refer the reader to my [Fl 09b] for the detailed analysis. Briefly the explanation is that the non-zero velocity eigenstates (and the boosts of 3-momentum eigenstates occurring in (6.3)) have non-instantaneous no-decay hyperplanes and the amplitude defined by (6.3) *is* the survival amplitude for finding the parent quanton on hyperplanes *parallel* to the no-decay hyperplane, but expressed in terms of the *time* interval between the hyperplanes instead of the *time-like* interval, orthogonal to the hyperplanes.

A time-like interval between parallel, space-like hyperplanes is always maximal in the direction orthogonal to the hyperplanes (the direction parallel to $\eta^{\mu}$) and this fact is the source of the *time* interval between parallel, non-instantaneous hyperplanes being shorter than the orthogonal time-like interval. While never being explicit on the matter, the discussion presented by Shirokov and Hegerfeldt reads as if the velocity eigenstates they work with are being regarded as having a definite no-decay *time*, $t = 0$. This motivates the misleading parameterization of the amplitude in terms of the *time*.

In the final analysis, there is no possibility of choosing between 3-momentum or velocity eigenstates as providing the most appropriate basis for UQs. Both are simply special cases of the generalized, space-like momentum eignstates, $| p, m; \eta, \tau >$, introduced in the previous section and all of which are required to adequately express the range of manifestation of UQs with arbitrary no-decay hyperplanes. The explicit declaration of the special cases in question is given by,

$$| \vec{k}, m; t > = | p = (0, \vec{k}), m; \eta = (1, \vec{0}), ct >, \tag{6.6a}$$

and

$$| \vec{u}, m > = | p = 0, m; \eta, 0 >, \tag{6.6b}$$

where, $\eta = (1, \vec{u}) / \sqrt{1 - u^2}$, is the 4-velocity corresponding to the 3-velocity, $\vec{u}$.

We now return to our central concern which is the analysis of the probability for finding the UQ on an *instantaneous* hyperplane when the no-decay hyperplane of that UQ is not merely distinct from, but *intersects* the first hyperplane. In such a case, in every frame, at least one of the two hyperplanes is non-instantaneous and thus, reverting to the general formalism in which arbitrary hyperplanes are considered is in order.

Accordingly, we generalize the analysis to emphasize that the quantity we are interested in is a special case of matrix elements of the form,

$$< p_2,m_2;\eta,\tau \,|\, \hat{\Pi}(\eta',\tau') \,|\, p_1,m_1;\eta,\tau >, \qquad (6.7)$$

where $\eta' \neq \eta$ and the projection operator, $\hat{\Pi}(\eta',\tau')$, is defined by,

$$\hat{\Pi}(\eta',\tau') = \sum_{m=-s}^{s} \int d_{\eta'}^3 p\, |\, p,m;\eta',\tau' >< p,m;\eta',\tau' \,| \;, \qquad (6.8)$$

and the projection operator, (2.14), is the special case, $\hat{\Pi}(t) = \hat{\Pi}(\eta_{(0)}, ct)$.

The surprise for the matrix elements, (6.7), is that their dependence on $\tau$ and $\tau'$ (when $\eta' \neq \eta$) is confined to an exactly calculable phase factor (**App. 6**), i.e.,

$$< p_2,m_2;\eta,\tau \,|\, \hat{\Pi}(\eta',\tau') \,|\, p_1,m_1;\eta,\tau > =$$

$$\exp\!\left[ (i/\hbar)\, \frac{\eta'(p_1-p_2)(\tau'-(\eta\eta')\tau)}{((\eta\eta')^2-1)} \right] < p_2,m_2;\eta,0\,|\hat{\Pi}(\eta',0)\,|\, p_1,m_1;\eta,0 >, \qquad (6.9)$$

In fact, if the matrix element was proportional to the three dimensional delta function of the momentum difference, $\delta_\eta^3(p_2-p_1)$, the phase factor would be completely superfluous and there would be no "time" dependence at all. But the matrix element is only proportional to the two dimensional delta function, $\delta_{\eta\eta'}^2(p_2-p_1)$, conserving the components of the momenta eigenvalues orthogonal to *both* $\eta$ and $\eta'$. The component parallel to $\eta'$ is not conserved. Nevertheless, the *diagonal* matrix element, with $\eta'(p_2-p_1)=0$, is "time" independent. It would seem that there is no *decay* at all with "time"! How are we to understand this?

In fact, an intuitive grasp of this result is not hard to come by. It stems from the fact that the SP spin-momentum eigenstates appearing in the matrix element are *not preferentially localized anywhere* on the $(\eta, \tau)$ no-decay hyperplane. As a consequence, the probability for finding the undecayed parent quanton on an intersecting hyperplane can depend on the hyperplanes only through their relationships with the momentum-spin eigenvalues and each other. In particular, this allows for a dependence on the '*angle of intersection*' of the hyperplanes, determined by $\eta$ and $\eta'$, but not on *where* the intersection occurs, i.e., on $\tau$ or $\tau'$ (**Fig. 1a**).

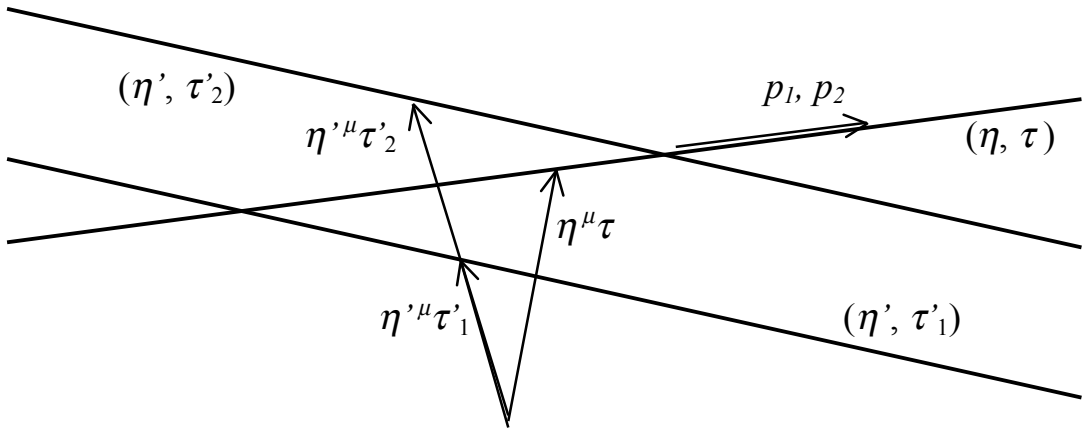

**Fig. 1a:** For SP momentum eigenstates with no-decay hyperplane, $(\eta, \tau)$, the probability (density) for survival on intersecting hyperplanes, $(\eta', \tau'_1)$ or $(\eta', \tau'_2)$ can not depend on the "time" variables, $\tau$ and $\tau'$. The reason is the absence of preferential localization of the momentum eigenstates on $(\eta, \tau)$ and the intrinsic relation between the intersecting hyperplanes, themselves, depending only on $\eta$ and $\eta'$.

Applying this result to (6.2) we have all the $t$ dependence confined to an off diagonal phase factor. *To obtain a time dependent survival probability for a boosted SP state we must have some degree of localization of the unstable parent on the $(\eta, 0)$ hyperplane!* (**Fig. 1b**)

We now turn to the calculation of the survival probability for a boosted unit norm state. As above, we actually calculate the more general expectation value, $<\psi; \eta, \tau \mid \hat{\Pi}(\eta', \tau') \mid \psi; \eta, \tau >$, to be able to see our survival probability, (6.1), as a special case. Unlike the above, however, the calculation will both probe more deeply and remain incomplete. First we have the boosted version of (3.1),

$$| p,m;\eta,\tau > = \int d^4q \, | q,m > \delta_\eta^3(p-q)\sqrt{2\eta q}\, r(q)\exp[(i/\hbar)\eta q\tau]\,, \qquad (6.10)$$

which will be employed in the analysis of the expectation value,

$$< \psi;\eta,\tau \,|\, \hat{\Pi}(\eta',\tau') \,|\, \psi;\eta,\tau >$$

$$= \sum_{m_2,m_1=-s}^{s} \int d_\eta^3 p_2 \, d_\eta^3 p_1 \, \psi_{m_2}(p_2)^* < p_2,m_2;\eta,\tau \,|\, \hat{\Pi}(\eta',\tau') \,|\, p_1,m_1;\eta,\tau > \psi_{m_1}(p_1)\,. \qquad (6.11)$$

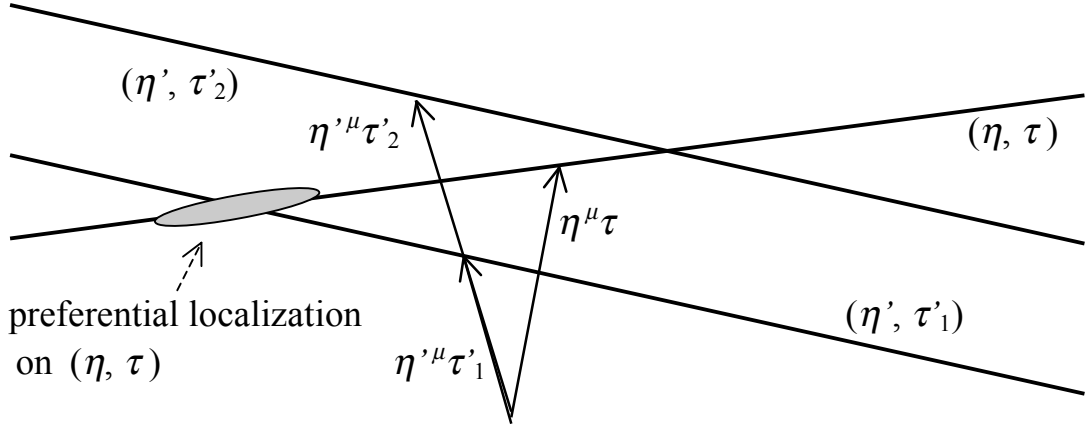

**Fig. 1b:** With the indicated preferential localization of the parent quanton on the no-decay hyperplane, $(\eta, \tau)$, the probability of finding the undecayed parent on the intersecting hyperplane, $(\eta', \tau'_1)$, will be higher than on, $(\eta', \tau'_2)$, at which more probability for decay will have accumulated.

On the right hand side we substitute (6.9) followed by (using (6.10)),

$$< p_2,m_2;\eta,0 \,|\, \hat{\Pi}(\eta',0) \,|\, p_1,m_1;\eta,0 >$$

$$= \int d^4q_2 \, d^4q_1 \, \delta_\eta^3(p_2-q_2)\sqrt{2\eta q_2}\, r^*(q_2) < q_2,m_2 \,|\, \hat{\Pi}(\eta',0) \,|\, q_1,m_1 > r(q_1)\sqrt{2\eta q_1}\, \delta_\eta^3(q_1-p_1)$$

$$= 2\,\delta_{\eta,\eta'}^2(p_2-p_1)\int d\omega_2 \, d\omega_1 \sqrt{\omega_2\,\omega_1}\, r^*(p_2+\eta\omega_2) r(p_1+\eta\omega_1)$$

$$< p_2 + \eta \omega_2, m_2 \mid \hat{\Pi}(\eta',0) \mid p_1 + \eta \omega_1, m_1 > . \qquad (6.12)$$

Next we have (using (6.8) and (6.10)),

$$< q_2, m_2 \mid \hat{\Pi}(\eta',0) \mid q_1, m_1 >$$

$$= \sum_{m'=-s}^{s} \int d_{\eta'}^3 p' < q_2, m_2 \mid p', m'; \eta', 0 >< p', m'; \eta', 0 \mid q_1, m_1 >$$

$$= \sum_{m'=-s}^{s} \int d_{\eta'}^3 p' \, \delta_{\eta'}^3 (q_2 - p') \delta_{m_2, m'} \sqrt{2 \eta' q_2} \, r(q_2) \, r*(q_1) \sqrt{2 \eta' q_1} \, \delta_{m', m_1} \delta_{\eta'}^3 (p' - q_1)$$

$$= \delta_{m_2, m_1} \delta_{\eta'}^3 (q_2 - q_1) 2 \sqrt{(\eta' q_2)(\eta' q_1)} \, r(q_2) \, r*(q_1) , \qquad (6.13)$$

where we have chosen to diagonalize all spin states along a single space-like direction parallel to the 2-manifold that comprises the intersection of the $(\eta, \tau)$ and $(\eta', \tau')$ hyperplanes.

A usefully compact expression for (6.12) (tolerating slight notational license) can now be obtained by introducing the conventions,

$$p_{1,2} + \eta \omega_{1,2} \equiv q_{1,2} \equiv q_{1 \eta, 2 \eta} + \eta (\eta q_{1,2}) , \qquad (6.14)$$

and employing them throughout (6.12). The combination of (6.12) and (6.13) then immediately yields,

$$< p_2, m_2; \eta, 0 \mid \hat{\Pi}(\eta', 0) \mid p_1, m_1; \eta, 0 > \equiv < q_{2\eta}, m_2; \eta, 0 \mid \hat{\Pi}(\eta', 0) \mid q_{1\eta}, m_1; \eta, 0 >$$

$$= \delta_{m_2, m_1} \int d \eta q_2 \, d \eta q_1 \, \delta_{\eta'}^3 (q_2 - q_1) \sqrt{(\eta q_2)(\eta' q_2)(\eta q_1)(\eta' q_1)} \, \frac{\sigma \left( \sqrt{q_2^2} \right)}{\sqrt{q_2^2}} \, \frac{\sigma \left( \sqrt{q_1^2} \right)}{\sqrt{q_1^2}} , \qquad (6.15)$$

where we have also employed (3.2). A more explicit rendering of (6.15) is provided in **App. 7**, where we find,

$$< p_2, m_2; \eta, 0 \mid \hat{\Pi}(\eta', 0) \mid p_1, m_1; \eta, 0 > =$$

$$= \frac{\delta_{m_2,m_1}\,\delta^2_{\eta',\eta}(p_2-p_1)}{\sqrt{(\eta\eta')^2-1}}\int d\omega_2\,d\omega_1\,\delta\!\left(\omega_1-\omega_2+\frac{(\eta\eta')\eta'(p_1-p_2)}{(\eta\eta')^2-1}\right)$$

$$\sqrt{\omega_2\omega_1\,\eta'(p_2+\eta\omega_2)\,\eta'(p_1+\eta\omega_1)}\;\frac{\sigma\!\left(\sqrt{p_2^2+\omega_2^2}\right)}{\sqrt{p_2^2+\omega_2^2}}\,\frac{\sigma\!\left(\sqrt{p_1^2+\omega_1^2}\right)}{\sqrt{p_1^2+\omega_1^2}}\,. \qquad (6.16)$$

We immediately notice that upon substitution of (6.16) into (6.9) and the moving of the "time" dependent phase factor under the influence of the integral, the 2nd delta function in (6.16) allows the replacement of the phase factor with,

$$\exp\!\left[(i/\hbar)\,\frac{\eta'(p_1-p_2)(\tau'-(\eta\eta')\tau)}{((\eta\eta')^2-1)}\right] \;=\; \exp\!\left[(i/\hbar)\frac{\omega_2-\omega_1}{\eta\eta'}(\tau'-(\eta\eta')\tau)\right]$$

$$= \exp\{(i/\hbar)(\eta q_1-\eta q_2)[\tau-(\tau'/\eta\eta')]\}\,. \qquad (6.17)$$

This is significant for getting the "energy-like", $\eta q_{1,2}$ quantities back into the "time" dependent exponent analogous to the case of (3.7). In particular, as we now turn to the calculation of the unit norm state expectation value, (6.11), we can exploit our notational conventions, (6.14), to write,

$$d_\eta^3 p_{1,2}\,d\eta q_{1,2}\equiv d_\eta^3 q_{1\eta,2\eta}\,d\eta q_{1,2}=d_\eta^3 q_{1,2}\,d\eta q_{1,2}=d^4 q_{1,2}=d_\eta^3 q_{1,2}\,d\eta' q_{1,2}\,, \qquad (6.18a)$$

and
$$\psi_{m_{1,2}}(p_{1,2})\equiv \psi_{\eta;m_{1,2}}(q_{1\eta,2\eta})\equiv \psi_{\eta;m_{1,2}}(q_{1,2})\,, \qquad (6.18b)$$

and then combine (6.18, 17, 15) and (6.9) to obtain for (6.11),

$$<\psi;\eta,\tau\,|\,\hat{\Pi}(\eta',\tau')\,|\,\psi;\eta,\tau> \;=$$

$$= \sum_{m=-s}^{s}\int d_\eta^3 q\left|\int d\eta' q\,\sqrt{(\eta q)(\eta' q)}\,\frac{\sigma\!\left(\sqrt{q^2}\right)}{\sqrt{q^2}}\,\psi_{\eta;m}(q)\exp\!\left[\frac{i}{\hbar}\eta q[\tau-(\tau'/\eta\eta')]\right]\right|^2. \qquad (6.19a)$$

Noticing that under the absolute value sign in (6.19a) the quantity, $q_{\eta'}$, is fixed and, therefore, its contribution to the phase factor can be cancelled and the replacement, $\eta q \equiv \eta q_{\eta'} + (\eta\eta')\eta' q \to (\eta\eta')\eta' q$, can be made in the phase factor, yielding,

$$< \psi; \eta, \tau \mid \hat{\Pi}(\eta', \tau') \mid \psi; \eta, \tau > =$$

$$\sum_{m=-s}^{s} \int d_\eta^3 q \left| \int d\eta' q \sqrt{(\eta q)(\eta' q)} \frac{\sigma\left(\sqrt{q^2}\right)}{\sqrt{q^2}} \psi_{\eta;m}(q) \exp\left[ -\frac{i}{\hbar} \eta' q (\tau' - (\eta\eta')\tau) \right] \right|^2. \quad (6.19b)$$

This relationship, for the special case of, $\eta' = \eta_{(0)} = (1, \vec{0})$, and, $\tau = 0$, but without recognizing the roles played by the intersecting survival and no-decay hyperplanes and neglecting spin, was expressed in non-covariant notation by Stefanovich [St 08]. In particular, Stefanovich noted the difficulty in having any value of $\tau'$ yield the value, unity, for the survival probability. He interpreted this as indicating that decays can be "caused by boosts" since $\mid \psi; \eta, 0 >$ is a boosted state, i.e., $\eta \neq \eta_{(0)}$. From our present perspective the failure of (6.19) to reach unity is simply due to the uniqueness of the no-decay hyperplane for an UQ. When searching for the UQ on a hyperplane other than the no-decay hyperplane, finding it, as opposed to finding decay products or formation precursors, will always be less than certain. It seems just as inappropriate to regard decays to be *caused* by boosts as it would be to regard decays to be *caused* by time translations. If cause is to be assigned, it is to the operation of interactions in the course of dynamical evolution from the no-decay hyperplane to *any* other hyperplane, including intersecting ones.

## 7. The lifetime of boosted ISP states

We now turn to our final task of integrating (6.19) over $\tau'$ to obtain the 'lifetime' associated with this generalized survival probability. But a conundrum confronts us immediately. Where do we *start* the integration? Unlike our first lifetime calculation, (4.1), where the no-decay hyperplane and the 'survival' hyperplane were parallel and the decay process obviously began at $t = 0$, here the 'survival' hyperplanes all intersect the no-decay hyperplane and no choice is the obvious beginning of the decay process. Presumably there is one among the survival hyperplanes which maximizes the survival probability; but which one?! From (6.19) a general determination does not look simple. On the other hand, **Fig. 1b** suggests that maximization will occur when the preferential localization region of the UQ on its no-decay hyperplane overlaps the intersection with the appropriate survival hyperplane. One could make this conjecture precise by considering the expectation value of a global, generalized position operator; the generalized Newton-Wigner operator, say, and demanding that it also lie on the *maximizing* survival hyperplane, $(\eta', \tau'_0)$, as well as on the no-decay hyperplane, $(\eta, \tau)$ i.e.,

$$\eta'_\mu < \psi; \eta, \tau \mid \hat{X}^\mu(\eta, \tau) \mid \psi; \eta, \tau > = \tau'_0, \qquad (7.1)$$

where, $\eta_\mu \hat{X}^\mu(\eta, \tau) = \tau$. But even if this plausible, but far from certain conjecture were accepted, the determination of $(\eta', \tau'_0)$ remains elusive.

Another consideration is motivated by the recognition that in the case of the lifetime, (4.1b), of an ISP state, the time integration from $-\infty$ to $+\infty$ would just double the result due to the decay being an inverse recapitulation of the formation of the UQ state from $-\infty$ to 0. In the ISP case this is easy to see due to the time reversal invariance of the probability, (3.7).

Does the same thing happen in the more general SP state case, thereby allowing the lifetime calculation to be just half the integral over $\tau'$ from $-\infty$ to $+\infty$ ? Again this seems plausible, but not so easy to demonstrate from (6.19). Some insight may be gained by examining the $\tau'$ integral of just the exponential factor in (6.19), understanding the expressions in the sense of generalized functions. We have, where, $\Delta = \eta' q_2 - \eta' q_1$ ,

$$\int_{\tau'_0}^{\infty} d\tau' \exp\left[(i/\hbar)\Delta(\tau' - (\eta\eta')\tau)\right] = \exp\left[(i/\hbar)\Delta(\tau'_0 - (\eta\eta')\tau)\right] i\hbar \left\{\frac{\mathrm{Pr}}{\Delta} - i\pi\delta(\Delta)\right\}$$

$$= i\hbar \frac{\mathrm{Pr}}{\Delta} \exp\left[(i/\hbar)\Delta(\tau'_0 - (\eta\eta')\tau)\right] + \pi\hbar\delta(\Delta) . \qquad (7.2)$$

where Pr denotes principle value. Now the generalized function limits of the first term on the right hand side of (7.2), for $\tau'_0 \to \pm\infty$, are, $\mp\pi\hbar\delta(\Delta)$, respectively. And since the expectation value, (6.19), is always real, there must be an intermediate value of $\tau'_0$ that renders the contribution of that first right hand side term in (7.2) to the $\tau'$ integral of (6.19) from $\tau'_0$ to $+\infty$ equal to zero. This intermediate value of $\tau'_0$ seems a likely candidate for the value at which the lifetime integral should begin. Furthermore, while we don't know what that intermediate value of $\tau'_0$ is, we do know what the resulting lifetime integral will yield, *viz.* just the contribution from the delta function term in (7.2)! We will proceed on the assumption that all of our conjectures concerning $\tau'_0$ yield roughly compatible and possibly exactly equal values and that the delta function term in (7.2) is the correct contribution. This assumption then yields for the lifetime,

$$T_\psi(\eta,\eta') = \int\limits_{\tau'_0}^{\infty} d\tau' <\psi;\eta,\tau \mid \hat{\Pi}(\eta',\tau') \mid \psi;\eta,\tau > = \frac{1}{2}\int\limits_{-\infty}^{\infty} d\tau' <\psi;\eta,\tau \mid \hat{\Pi}(\eta',\tau') \mid \psi;\eta,\tau >$$

$$= \pi\hbar \int d_\eta^3 p \left[ \left( \sum_m |\psi_m(p)|^2 \right) \int d\omega\, \omega (\eta' p + (\eta\eta')\omega) \frac{\sigma^2\left(\sqrt{p^2+\omega^2}\right)}{p^2+\omega^2} \right], \qquad (7.3)$$

where the variables of integration go through the transitions from (6.19) to (7.3) indicated by the sequence,

$$d\tau' d_\eta^3 q\, d\eta' q_2\, d\eta' q_1 \rightarrow d_\eta^3 q\, d\eta' q_2\, d\eta' q_1\, \delta(\eta' q_2 - \eta' q_1) \rightarrow d_\eta^3 q\, d\eta' q$$

$$= d^4 q = d_\eta^3 q\, d\eta q = d_\eta^3 p\, d\omega \qquad (7.4)$$

Finally, we reintroduce the mass spectrum variable, $\mu = \sqrt{p^2+\omega^2}$ . Noting that, $d_\eta^3 p\, d\omega = d_\eta^3 p\, d\mu\,\mu$ , we can rewrite (7.3) in the form,

$$T_\psi(\eta,\eta') = \pi\hbar \int d_\eta^3 p \left[ \left( \sum_m |\psi_m(p)|^2 \right) \int\limits_{\mu_{min}}^{\infty} d\mu\, \sigma^2(\mu) \frac{\sqrt{\mu^2-p^2}}{\mu} \left( \eta\eta' + \frac{\eta' p}{\sqrt{\mu^2-p^2}} \right) \right]. \quad (7.5)$$

This is to be compared first to (4.4), to which (7.5) reduces if $\eta' = \eta = \eta^{(0)} = (1,\vec{0})$ . But what of the more interesting case for which we actually obtained (7.5), i.e., the case where $\eta' = \eta^{(0)} = (1,\vec{0})$ while $\eta$ remains arbitrary? This is the case of (6.1) for which we require the substitutions (in (7.5)),

$$\psi_m(p) \rightarrow \psi_{B,m}(p) = \psi_{B,m}(B(\vec{u})(0,\vec{k})) = \psi_m(\vec{k}) , \qquad (7.6a)$$

$$d_\eta^3 p = d^3k, \quad \sqrt{\mu^2 - p^2} = \sqrt{\mu^2 + k^2}, \quad \eta\eta' \to \eta\eta_{(0)} = \eta^0 = \frac{1}{\sqrt{1-u^2}}, \qquad (7.6b)$$

and

$$\eta' p \to \eta_{(0)} p = p^0 = (B(\vec{u})(0,\vec{k}))^0 = \frac{\vec{u}\cdot\vec{k}}{\sqrt{1-u^2}}. \qquad (7.6c)$$

With these substitutions and employing the notation,

$$T_k(\mu) \equiv \pi\hbar\,\sigma(\mu)\,\frac{\sqrt{\mu^2+k^2}}{\mu}, \qquad (7.7)$$

(7.5) can be expressed in the form,

$$T_{\psi_B}(\eta,\eta^{(0)}) = \int d^3k \left[ \sum_m \left|\psi_m(\vec{k})\right|^2 \int_{\mu_{\min}}^{\infty} d\mu\,\sigma(\mu)\,\frac{T_k(\mu)}{\sqrt{1-u^2}}\left(1+\frac{\vec{u}\cdot\vec{k}}{\sqrt{\mu^2+k^2}}\right) \right], \qquad (7.8)$$

We have chosen this particular form for (7.8) to facilitate comparison with the classical equation, (1.5), for the dilation of a lifetime of an unstable particle with an initial non-zero momentum. We repeat that expression here.

$$T_w = \frac{T_0}{\sqrt{1-w^2}} = \frac{T_v}{\sqrt{1-u^2}}[1+\vec{u}\cdot\vec{v}] = \frac{T_k}{\sqrt{1-u^2}}\left[1+\frac{\vec{u}\cdot\vec{k}}{\sqrt{m^2+k^2}}\right]. \qquad (1.5)$$

Clearly, the quantum lifetime dilation differs, formally, from the classical case in a completely plausible way, i.e., by summing the classical lifetime dilation dependence over the spectra of rest mass, momenta and spin contributions that contribute to the initial state. What could be more natural?! But we must not lose sight of the serious difference in the relation between the momentum contribution and the boost contribution in the quantum and classical cases. The classical particle can reach the composite velocity, $\vec{w}$, *either* from the boost by velocity, $\vec{u}$, from a state with momentum, $\vec{k}$, or via a boost with velocity, $\vec{w}$, from a state with zero momentum or velocity. Because of the non-trivial mass spectrum, however, the 3-momentum composition of the UQ can not be acquired through boosts of that quanton from zero 3-momentum. The only dependence of the quanton lifetime (7.8) on a boost of the quanton is the dependence on $\vec{u}$, not the dependence on $\vec{k}$. The 3-momentum composition of the UQ state in the frame of reference in which the no-decay hyperplane is instantaneous and the boost of that state to a frame in which the no-decay hyperplane is not instantaneous comprise a *two-fold origin* of the dilation of the UQ lifetime from its minimum possible value.

We will close this section with the analogue of (4.6) having the same relation to (7.5) that (4.6) has to (4.4). Recalling the derivation of (4.6) and by simple inspection of (7.5) we see that,

$$T_\psi(\eta, \eta') = \pi\hbar < \psi; \eta, \tau \mid \sigma\left(\sqrt{\hat{P}^2}\right)\eta'\hat{P} / \sqrt{\hat{P}^2} \mid \psi; \eta, \tau >, \qquad (7.9)$$

independent of $\tau$. Upon setting $\eta' = \eta_{(0)}$, and writing,

$$\mid \psi; \eta, 0 > = \hat{U}(B(\vec{u})) \mid \psi >, \qquad (7.10)$$

we find,

$$T_\psi(\eta, \eta_{(0)}) = \pi\hbar < \psi \mid \hat{U}^\dagger(B(\vec{u}))\sigma\left(\sqrt{\hat{P}^2}\right)\frac{\hat{P}^0}{\sqrt{\hat{P}^2}}\hat{U}(B(\vec{u})) \mid \psi >$$

$$= \pi\hbar < \psi \mid \frac{\sigma\left(\sqrt{\hat{P}^2}\right)}{\sqrt{\hat{P}^2}}\frac{\hat{P}^0 + \vec{u}\cdot\hat{\vec{P}}}{\sqrt{1-u^2}} \mid \psi >, \qquad (7.11)$$

as the corresponding version of (7.8). In particular, if $\mid \psi >$, with instantaneous no-decay hyperplane, also has $< \hat{\vec{P}} >_\psi = \vec{0}$, we then have,

$$T_{\hat{U}(B(\vec{u}))|\psi>} = \pi\hbar < \psi \mid \sigma\left(\sqrt{\hat{P}^2}\right)\hat{P}^0 / \sqrt{\hat{P}^2} \mid \psi > / \sqrt{1-u^2} = T_{|\psi>} / \sqrt{1-u^2}, \qquad (7.12)$$

an instance of *exact* Einstein dilation of a quanton lifetime which, we note in passing, includes arbitrarily close approximations to velocity eigenstates!

## Appendix 1: Spin operator for ISP states

Beginning with (2.9), our (A1.1), we have,

$$\hat{S}^3 \mid \vec{k}, m; t > = \mid \vec{k}, m; t > \hbar m, \qquad (A1.1)$$

and

$$(\hat{S}^1 \pm i\hat{S}^2)\,|\,\vec{k},m;t> = |\,\vec{k},m\pm 1;t> \hbar\sqrt{(s\mp m)(s\pm m+1)} \quad . \tag{A1.2}$$

Finally, for $\hat{\vec{J}}$ to be the generator of infinitesimal rotations, we must have,

$$\hat{\vec{J}}\,|\,\vec{k},m;t> = i\hbar(\vec{k}\times\partial/\partial\vec{k})\,|\,\vec{k},m;t> + \hat{\vec{S}}\,|\,\vec{k},m;t> , \tag{A1.3}$$

which yields,

$$\exp[-(i/\hbar)\hat{\vec{J}}\cdot\vec{n}\theta]\,|\,\vec{k},m;t> = \sum_{m'=-s}^{s}|\,R(\vec{n}\theta)\vec{k},m';t> Y_{m',m}^{s}(\vec{n}\theta), \tag{A1.4}$$

where,

$$R(\vec{n}\theta)\vec{k} = \vec{k}_{\|} + \vec{k}_{\perp}\cos\theta + \vec{n}\times\vec{k}\sin\theta . \tag{A1.5}$$

is the rotated momentum and the subscripts $\perp$ and $\|$ denote components perpendicular and parallel, respectively, to the rotation axis, $\vec{n}$.

## Appendix 2: 3-momentum eigenstates as sums over 4-momentum eigenstates

We begin with,

$$|\,\vec{k},m;t> = \int dq^0 \delta(\hat{P}^0 - q^0)\,|\,\vec{k},m;t> := \int dq^0\,\delta(\hat{P}^0 - q^0)\exp[(i/\hbar)\hat{P}^0 t]\,|\,\vec{k},m;0>$$

$$= \int d\,q^0\,|\,q^0,\vec{k},m> f_m(q^0,\vec{k})\,\exp[(i/\hbar)\,q^0 t]$$

$$= \int d^4 q\,|\,q,m> \delta^3(\vec{q}-\vec{k})\,f_m(q)\,\exp[(i/\hbar)\,q^0 t]. \tag{A2.1}$$

Now assuming the normalization,

$$<q',m'\,|\,q,m> = \delta^4(q'-q)\delta_{m'm} \quad , \tag{A2.2}$$

we must have,

$$<\vec{k}_2,m_2;t\,|\delta(\hat{P}^2 - \mu^2)\,|\,\vec{k}_1,m_1;t>$$

$$= \int d^4 q\,f_{m_2}(q)*\delta^3(\vec{q}-\vec{k}_2)\delta_{m_2,m_1}\delta(q^2 - \mu^2)\,\delta^3(\vec{q}-\vec{k}_1)f_{m_1}(q)$$

$$= \delta_{m_2,m_1}\delta^3(\vec{k}_2-\vec{k}_1)\int dq^0\,\delta((q^0)^2 - k_1^2 - \mu^2)\,|\,f_{m_1}((q^0,\vec{k}_1))\,|^2$$

$$= \delta_{m_2, m_1} \delta^3(\vec{k}_2 - \vec{k}_1) \int dq^0 \frac{1}{2q^0} \delta(q^0 - \sqrt{k_1^2 + \mu^2}) \mid f_{m_1}((q^0, \vec{k}_1)) \mid^2$$

$$= \delta_{m_2, m_1} \delta^3(\vec{k}_2 - \vec{k}_1) \frac{1}{2\sqrt{k_1^2 + \mu^2}} \left| f_{m_1}\left(\left(\sqrt{k_1^2 + \mu^2}, \vec{k}_1\right)\right) \right|^2. \qquad (A2.3)$$

But according to (2.11) we must have,

$$< \vec{k}_2, m_2; t \mid \delta(\hat{P}^2 - \mu^2) \mid \vec{k}_1, m_1; t > = \delta^3(\vec{k}_2 - \vec{k}_1) \delta_{m_2, m_1} \sigma(\mu) / 2\mu, \qquad (A2.4)$$

Therefore we can write,

$$f_m(q) = \sqrt{2q^0} \, r(q) \ , \qquad (A2.5)$$

where,

$$\mid r(q) \mid^2 = \sigma\left(\sqrt{q^2}\right) / 2\sqrt{q^2} \ . \qquad (A2.6)$$

## Appendix 3: Calculation of the ISP lifetime, (4.3)

The calculation is straightforward starting from the left hand side of (4.2). Thus,

$$T_k = \int_0^\infty dt \left| \int_{\mu_{\min}}^\infty d\mu \, \sigma(\mu) \exp\left[-(i/\hbar)\sqrt{\mu^2 + k^2} \, t\right] \right|^2$$

$$= \lim_{\varepsilon \to 0+} \int_{\mu_{\min}}^\infty d\mu' d\mu \, \sigma(\mu') \sigma(\mu) \left[\int_0^\infty dt \exp\left[(i/\hbar)\left(\sqrt{\mu'^2 + k^2} - \sqrt{\mu^2 + k^2} + i\varepsilon\right) t\right]\right]$$

$$= \lim_{\varepsilon \to 0+} \int_{\mu_{\min}}^\infty d\mu' d\mu \, \sigma(\mu') \sigma(\mu) \frac{i\hbar}{\sqrt{\mu'^2 + k^2} - \sqrt{\mu^2 + k^2} + i\varepsilon}$$

$$= \int_{\mu_{\min}}^\infty d\mu' d\mu \, \sigma(\mu') \sigma(\mu) \, \hbar \left[-\frac{i \, Pv}{\left(\sqrt{\mu'^2 + k^2} - \sqrt{\mu^2 + k^2}\right)} + \pi \delta\left(\sqrt{\mu'^2 + k^2} - \sqrt{\mu^2 + k^2}\right)\right]$$

$$= \pi \hbar \int_{\mu_{\min}}^\infty d\mu' d\mu \, \sigma(\mu') \sigma(\mu) \, \delta\left(\sqrt{\mu'^2 + k^2} - \sqrt{\mu^2 + k^2}\right)$$

$$= \pi\hbar \int\limits_{\mu_{min}}^{\infty} d\mu' d\mu \, \sigma(\mu') \sigma(\mu) \left( \sqrt{\mu^2 + k^2} \, / \, \mu \right) \delta(\mu' - \mu)$$

$$= \pi\hbar \int\limits_{\mu_{min}}^{\infty} d\mu \, \sigma(\mu)^2 \left( \sqrt{\mu^2 + k^2} \, / \, \mu \right), \tag{A3.1}$$

as claimed.

## Appendix 4: Spin operators for boosted SP states

Starting with (5.5),

$$\hat{U}^{\dagger}(B(\vec{u})) \hat{S}^{\mu}(\eta) \hat{U}(B(\vec{u})) = \hat{U}^{\dagger}(B(\vec{u})) (\hat{S}^0(\eta), \hat{\vec{S}}(\eta)) \hat{U}(B(\vec{u}))$$

$$= B(\vec{u}) (\hat{S}^0((1,\vec{0})), \hat{\vec{S}}((1,\vec{0}))) = B(\vec{u})(0, \hat{\vec{S}}) = \left( \frac{\vec{u} \cdot \hat{\vec{S}}}{\sqrt{1-u^2}}, \hat{\vec{S}}_{\perp} + \frac{\hat{\vec{S}}_{\parallel}}{\sqrt{1-u^2}} \right), \tag{A4.1}$$

we have,

$$\hat{U}^{\dagger}(B(\vec{u})) (\hat{S}^0(\eta), \hat{\vec{S}}(\eta)) \hat{U}(B(\vec{u})) = \left( \frac{\vec{u} \cdot \hat{\vec{S}}}{\sqrt{1-u^2}}, \hat{\vec{S}}_{\perp} + \frac{\hat{\vec{S}}_{\parallel}}{\sqrt{1-u^2}} \right)$$

$$= \left( \vec{\eta} \cdot \hat{\vec{S}}, \hat{\vec{S}} - \frac{\vec{\eta}(\vec{\eta} \cdot \hat{\vec{S}})}{\vec{\eta}^2} + \eta^0 \frac{\vec{\eta}(\vec{\eta} \cdot \hat{\vec{S}})}{\vec{\eta}^2} \right) = \left( \vec{\eta} \cdot \hat{\vec{S}}, \hat{\vec{S}} + \frac{\vec{\eta}(\vec{\eta} \cdot \hat{\vec{S}})}{\eta^0 + 1} \right). \tag{A4.2}$$

Consequently,

$$\hat{U}^{\dagger}(B(\vec{u})) \hat{\vec{S}}(\eta) \hat{U}(B(\vec{u})) = \hat{\vec{S}} + \frac{\vec{\eta}}{\eta^0 + 1} \hat{U}^{\dagger}(B(\vec{u})) \hat{S}^0(\eta) \hat{U}(B(\vec{u})),$$

or,

$$\hat{U}^{\dagger}(B(\vec{u})) \hat{\vec{S}}(\eta) \hat{U}(B(\vec{u})) - \frac{\vec{\eta}}{\eta^0 + 1} \hat{U}^{\dagger}(B(\vec{u})) \hat{S}^0(\eta) \hat{U}(B(\vec{u})) = \hat{\vec{S}},$$

or,

$$\left[ \hat{\vec{S}}(\eta) - \frac{\vec{\eta}}{\eta^0 + 1} \hat{S}^0(\eta) \right] \hat{U}(B(\vec{u})) = \hat{U}(B(\vec{u})) \hat{\vec{S}}, \tag{A4.3}$$

as claimed.

## Appendix 5: The consistency of (5.17)

From the relations, (5.14,15), and the definition,

$$\hat{K}^\mu(\eta) \equiv \hat{P}^\mu - \eta^\mu(\eta\hat{P}) \,, \qquad (A5.1)$$

the Lie algebra of the inhomogeneous Lorentz group takes the form,

$$[\hat{K}^\mu(\eta), \hat{K}^\nu(\eta)] = [\eta\hat{P}, \hat{K}^\nu(\eta)] = [\eta\hat{P}, \hat{J}^\nu(\eta)] = 0 \,, \qquad (A5.2a)$$

$$[\hat{J}^\mu(\eta), (\hat{J}^\nu(\eta), \hat{K}^\nu(\eta), \hat{N}^\nu(\eta))] = i\hbar\, \varepsilon^{\mu\nu\alpha\beta}(\hat{J}_\alpha(\eta), \hat{K}_\alpha(\eta), \hat{N}_\alpha(\eta))\eta_\beta \,, \qquad (A5.2b)$$

$$[\hat{N}^\mu(\eta), \hat{N}^\nu(\eta)] = -i\hbar\, \varepsilon^{\mu\nu\alpha\beta}\hat{J}_\alpha(\eta)\eta_\beta \,, \qquad (A5.2c)$$

$$[\hat{N}^\mu(\eta), \hat{K}^\nu(\eta)] = i\hbar(\eta^\mu\eta^\nu - \eta^{\mu\nu})\eta\hat{P} \,, \qquad (A5.2d)$$

and,

$$[\hat{N}^\mu(\eta), \eta\hat{P}] = i\hbar\, \hat{K}^\mu(\eta) \,. \qquad (A5.2e)$$

Introducing, $\hat{Q}^\mu(\eta,\tau) = \hat{X}^\mu(\eta,\tau) - \eta^\mu\tau$, the implicit definitions, (5.16), can be rewritten as,

$$\hat{J}^\mu(\eta) = -\,\varepsilon^{\mu\alpha\beta\gamma}\hat{Q}_\alpha(\eta,\tau)\hat{K}_\beta(\eta)\eta_\gamma + \hat{S}^\mu(\eta) \,, \qquad (A5.3a)$$

and,

$$\hat{N}^\mu(\eta) = \eta\hat{P}{:}\hat{Q}^\mu(\eta,\tau) - \tau\hat{K}^\mu(\eta) - \frac{\varepsilon^{\mu\alpha\beta\gamma}\hat{K}_\alpha(\eta)\hat{S}_\beta(\eta)\eta_\gamma}{\eta\hat{P} + \sqrt{\hat{P}^2}} \,, \qquad (A5.3b)$$

from which the preceeding commutation relations imply,

$$[\hat{Q}^\mu(\eta,\tau), \hat{Q}^\nu(\eta,\tau)] = [\hat{Q}^\mu(\eta,\tau), \hat{S}^\nu(\eta)] = [\hat{K}^\mu(\eta), \hat{S}^\nu(\eta)] = [\eta\hat{P}, \hat{S}^\nu(\eta)] = 0 \,, \qquad (A5.4a)$$

$$[\hat{Q}^\mu(\eta,\tau), \hat{K}^\nu(\eta)] = i\hbar(\eta^\mu\eta^\nu - \eta^{\mu\nu}) \,, \qquad (A5.4b)$$

$$[\hat{Q}^\mu(\eta,\tau), \eta\hat{P}] = i\hbar\, \hat{K}^\mu(\eta)/\eta\hat{P} \,, \qquad (A5.4c)$$

and,

$$[\hat{S}^\mu(\eta), \hat{S}^\nu(\eta)] = i\hbar\, \varepsilon^{\mu\nu\alpha\beta}\hat{S}_\alpha(\eta)\eta_\beta \,. \qquad (A5.4d)$$

From (A5.4a, b) it follows that,

$$\hat{S}^\mu(\eta)\exp[-(i/\hbar)\hat{Q}^\mu(\eta,\tau)\Delta_\mu] = \exp[-(i/\hbar)\hat{Q}^\mu(\eta,\tau)\Delta_\mu]\,\hat{S}^\mu(\eta),\qquad\text{(A5.5a)}$$

and,

$$\hat{K}^\mu(\eta)\exp[-(i/\hbar)\hat{Q}^\mu(\eta,\tau)\Delta_\mu] = \exp[-(i/\hbar)\hat{Q}^\mu(\eta,\tau)\Delta_\mu](\hat{K}^\mu(\eta)+\Delta^\mu),\quad\text{(A5.5b)}$$

where $\Delta^\mu$ satisfies, $\Delta\eta = 0$. This permits us to consistently assume,

$$\exp[-(i/\hbar)\hat{Q}^\mu(\eta,\tau)\Delta_\mu]\,|\,p,m;\eta,\tau> = |\,p+\Delta,m;\eta,\tau>,\qquad\text{(A5.6)}$$

which is the same as (5.17).

## Appendix 6: Derivation of (6.9)

We begin with,

$$\exp[(i/\hbar)\hat{P}\lambda]\,|\,p,m;\eta,\tau> = |\,p,m;\eta,\tau+\eta\lambda>\exp[(i/\hbar)p\lambda]\;,\qquad\text{(A6.1)}$$

and,

$$\exp[(i/\hbar)\hat{P}\lambda]\hat{\Pi}(\eta,\tau)\exp[-(i/\hbar)\hat{P}\lambda] = \hat{\Pi}(\eta,\tau+\eta\lambda),\qquad\text{(A6.2)}$$

for arbitrary 4-displacement, $\lambda^\mu$, which follow from the SP momentum eigenstates being eigenstates of space-like translations parallel to the no-decay hyperplane, $(\eta,\tau)$, and $\eta\hat{P}$ being the generator of time-like displacements normal to the no-decay hyperplane.

From (A6.1, 2) it immediately follows that,

$$<p_2,m_2;\eta,\tau\,|\,\hat{\Pi}(\eta',\tau')\,|\,p_1,m_1;\eta,\tau> =$$

$$<p_2,m_2;\eta,\tau+\eta\lambda\,|\,\hat{\Pi}(\eta',\tau'+\eta'\lambda)\,|\,p_1,m_1;\eta,\tau+\eta\lambda>\exp[(i/\hbar)(p_1-p_2)\lambda]\qquad\text{(A6.3)}$$

For any $\lambda^\mu$ that satisfies, $\eta\lambda = -\tau$ and $\eta'\lambda = -\tau'$, (A6.3) becomes,

$$<p_2,m_2;\eta,\tau\,|\,\hat{\Pi}(\eta',\tau')\,|\,p_1,m_1;\eta,\tau> =$$

$$<p_2,m_2;\eta,0\,|\,\hat{\Pi}(\eta',0)\,|\,p_1,m_1;\eta,0>\exp[(i/\hbar)(p_1-p_2)\lambda]\qquad\text{(A6.4)}$$

One such $\lambda^\mu$ is given by,

$$\lambda = \eta \frac{\tau - (\eta\eta')\tau'}{(\eta\eta')^2 - 1} + \eta' \frac{\tau' - (\eta\eta')\tau}{(\eta\eta')^2 - 1} \ , \tag{A6.5}$$

and the most general such $\lambda^\mu$ differs from (A6.5) by an arbitrary 4-vector orthogonal to both, $\eta$ and $\eta'$. This is consistent with (A6.4) only if the matrix element vanishes whenever the momentum difference, $p_1 - p_2$, is not parallel to the $(\eta, \eta')$ plane, as, indeed, we see to be the case in (6.12). Upon substituting (A6.5) into (A6.4), we obtain (6.9).

## Appendix 7: Derivation of (6.16)

Replacing, $q_{1,2}$ by $p_{1,2} + \eta\omega_{1,2}$, respectively, in (6.13), we have,

$$< p_2 + \eta\omega_2, m_2 \mid \hat\Pi(\eta', 0) \mid p_1 + \eta\omega_1, m_1 >$$

$$= \delta_{m_2, m_1} \delta_{\eta'}^3(p_2 - p_1 + \eta(\omega_2 - \omega_1)) 2\sqrt{\eta'(p_2 + \eta\omega_2)\eta'(p_1 + \eta\omega_1)}$$

$$r(p_2 + \eta\omega_2)r*(p_1 + \eta\omega_1). \tag{A7.1}$$

Substituting this into (6.12) we obtain

$$< p_2, m_2; \eta, 0 \mid \hat\Pi(\eta', 0) \mid p_1, m_1; \eta, 0 >$$

$$= 4 \delta_{m_2, m_1} \int d\omega_2 \, d\omega_1 \, \delta_{\eta'}^3(p_2 + \eta\omega_2 - p_1 + \eta\omega_1)$$

$$\sqrt{\omega_2\omega_1 \eta'(p_2 + \eta\omega_2)\eta'(p_1 + \eta\omega_1)} \mid r(p_2 + \eta\omega_2) \mid^2 \mid r(p_1 + \eta\omega_1) \mid^2$$

$$= 4 \delta_{m_2, m_1} \delta_{\eta', \eta}^2(p_2 - p_1) \int d\omega_2 \, d\omega_1 \, \delta(e'[p_2 + \eta\omega_2 - p_1 + \eta\omega_1])$$

$$\sqrt{\omega_2\omega_1 \eta'(p_2 + \eta\omega_2)\eta'(p_1 + \eta\omega_1)} \mid r(p_2 + \eta\omega_2) \mid^2 \mid r(p_1 + \eta\omega_1) \mid^2, \tag{A7.2}$$

where,

$$\delta_{\eta'}^3(p_2 + \eta\omega_2 - p_1 + \eta\omega_1) = \delta_{\eta', \eta}^2(p_2 - p_1)\delta(e'[p_2 + \eta\omega_2 - p_1 + \eta\omega_1]), \tag{A7.3}$$

and, $e' = (\eta - \eta'(\eta\eta')) / \sqrt{(\eta\eta')^2 - 1}$, is the space-like unit vector orthogonal to $\eta'$ and lying in the $(\eta, \eta')$ 2-plane. From,

$$\delta(e'[p_2 + \eta\omega_2 - p_1 + \eta\omega_1]) = \sqrt{(\eta\eta')^2 - 1}\, \delta((\eta\eta')\eta'(p_1 - p_2) + ((\eta\eta')^2 - 1)(\omega_1 - \omega_2))$$

$$= \frac{1}{\sqrt{(\eta\eta')^2 - 1}}\, \delta\left(\omega_1 - \omega_2 + \frac{(\eta\eta')\eta'(p_1 - p_2)}{(\eta\eta')^2 - 1}\right), \qquad (A7.4)$$

we obtain,

$$< p_2, m_2; \eta, 0 \,|\, \hat{\Pi}(\eta', 0) \,|\, p_1, m_1; \eta, 0 >$$

$$= \frac{4\, \delta_{m_2, m_1}\, \delta^2_{\eta', \eta}(p_2 - p_1)}{\sqrt{(\eta\eta')^2 - 1}} \int d\omega_2\, d\omega_1\, \delta\left(\omega_1 - \omega_2 + \frac{(\eta\eta')\eta'(p_1 - p_2)}{(\eta\eta')^2 - 1}\right)$$

$$\sqrt{\omega_2\omega_1}\, \eta'(p_2 + \eta\omega_2)\eta'(p_1 + \eta\omega_1)\, |\, r(p_2 + \eta\omega_2)\,|^2 |\, r(p_1 + \eta\omega_1)\,|^2. \qquad (A7.5)$$

This becomes (6.16) upon using (3.2).